\documentstyle{mn}
\input{psfig.tex}
\begin{document}
\def\ltsima{$\; \buildrel < \over \sim \;$}
\def\simlt{\lower.5ex\hbox{\ltsima}}
\def\gtsima{$\; \buildrel > \over \sim \;$}
\def\simgt{\lower.5ex\hbox{\gtsima}}

\title[The recent history of the X-ray absorber in NGC\thinspace3516]
{The recent history of the X-ray absorber in NGC\thinspace3516}

\author[M. Guainazzi et al.]
{M. Guainazzi$^{1,2}$, W. Marshall$^{2,3}$, A. N. Parmar$^2$ \\ ~ \\
$^1$ XMM-Newton SOC, VILSPA ESA, Apartado 50727, E-28080 Madrid, Spain \\
$^2$ Astrophysics Division, Space Science Department of
ESA, ESTEC, Postbus 299, NL-2200 AG Noordwijk, The Netherlands \\
$^3$ Leicester University, Leicester LE1 7RH, United Kingdom \\
}

\maketitle

\begin{abstract}

We present two BeppoSAX observations of the bright Seyfert 1 galaxy 
NGC\thinspace3516,
performed four months apart (late 1996/early 1997). The earlier spectrum is
considerably weaker and harder
in the whole 0.1--50~keV energy range.
In addition, the
warm absorber oxygen features, which were already observed with ROSAT
(Mathur et al. 1997) and ASCA (Kriss et al. 1996), are much less pronounced.
The most straightforward explanation is that in 1996 November
NGC\thinspace3516 was being seen through a substantial (${\rm
N_H \simeq 10^{22}}$~cm$^{-2}$) column of cold material.
This is the first confirmation with
modern instrumentation that NGC\thinspace3516 indeed
undergoes phases of strong cold X-ray absorption. We speculate that these
intervals may be connected to the onset of the remarkably variable
UV absorption system, making NGC\thinspace3516 the best known
example of low-luminosity Broad Absorption Line (BAL) quasar. 
The absorbing matter could be
provided by clouds ablated from the rim of the circumnuclear molecular
torus, seen at a rather high inclination angle.

\end{abstract}


\section{Introduction}

The X-ray spectra of type 1 Seyferts
do not show substantial amounts of photoelectric
absorption by neutral matter (Mushotzky 1982; Turner \& Pounds 1989). This is
consistent with a
scenario where these objects are seen along the axis of a dusty molecular
torus surrounding the nucleus (Antonucci \& Miller 1985; Antonucci
1993).
However, about 50\% of Seyfert~1 galaxies exhibit soft X-ray features due to
photoelectric absorption
by highly ionized matter, mostly He- and
H-like oxygen (Reynolds 1997; George et al. 1998a).
The physical condition of the
absorber and its location are still matter of debate.
The existence of a copious source of high-energy photons in 
Active Galactic Nuclei (AGN)
suggests that the ionization structure of the absorbing
material could be driven by the same non-thermal nuclear high-energy continuum
that we observe. However, in several cases (and, most noticeably, in the best
studied ones: MCG-6-30-15, Fabian et al. 1994; Reynolds et al. 1995;
NGC~4051, McHardy et al. 1995; Guainazzi et al. 1996),
the variability patterns on short
({\it i.e.}: hours) and medium ({\it i.e.}: days/weeks) timescales suggest
a more complex picture than a single-zone medium in thermal and ionization
equilibrium with the primary (nuclear)
continuum. A physically stratified medium (Otani
et al. 1996; George et al. 1998b) or
models of non-equilibrium and/or collisional excitation
(Nicastro et al. 1999a) are two
alternative possibilities. The high-resolution observations to
be provided by the grating systems on-board {\it Chandra} and XMM,
will allow the ionization structure of the absorbing regions to
be studied in detail.

However, variability studies are still an important tool
in helping determine the location of the absorbing material.
These studies generally
rely on sparse and mostly episodic monitoring
campaigns. The original
post-ROSAT suggestion that the warm absorber is coincident or co-located
with the Broad Line Region (Mushotzky et al. 1993)
is consistent with current observational evidence
(Guainazzi et al. 1996;
Fabian et al. 1994; George et al. 1998a),
and is discussed in detail for the case of NGC\thinspace3516 by
Kolman et al. (1993), who derive a loose constraint on the distance
to the absorber of 0.01--9~pc.

Recently, Boller et al. (1997) and Guainazzi et al. (1998)
have speculated that the dramatic
variability observed in a few Narrow Line Seyfert Galaxies 
may be due to the interposition of randomly
distributed ``bricks'' of material with 
${\rm N_H \simgt 10^{22}}$~cm$^{-2}$ along
the line of sight. The sampling of this variability and the
available spectral resolution is generally not good enough to exclude 
alternative mechanisms, and it is still not clear how general this
phenomenology is in Seyfert galaxies as a class.

This {\it paper} presents two recent BeppoSAX observations of the
nearby (${\rm z=0.0088}$) Seyfert~1 galaxy
NGC\thinspace3516, which reveal new interesting features, relevant to the
above issues. 
NGC\thinspace3516, being one of the brightest Seyfert 1s in the X-ray sky,
has been extensively studied by all the major X-ray observatories.
Low-resolution, bandwidth-limited observations in the 1980s 
suggested an extreme
variability, more pronounced in the soft X-ray band. Two observations with
the Imaging Proportional Counter on-board {\it Einstein} showed
an increase by a factor of 5 of the 0.2--4.0~keV flux, accompanied by
a large change in spectral shape (the lower the flux, the
flatter the spectrum; Kruper et al. 1990). Monitoring with the
Monitor Proportional Counter (MPC) on-board {\it Einstein} and with EXOSAT
(Ghosh \& Soundararajaperumal 1991)
revealed strong and highly variable
(0.5--2$\times 10^{23}$~cm$^{-3}$) photoelectric absorption. However,
with the advent of the moderate energy resolution provided by the ROSAT
Position Sensitive Proportional Counter
and by ASCA,
it has been possible to identify NGC\thinspace3516 as one of the 
``warm absorbed''
Seyfert~1s. Comparison of the ASCA observations with detailed photoionization
models suggests the presence of a complex multi-zone warm absorber, with
a broad ({\it i.e.} almost
one order of magnitude) range of ionization parameters and column densities
(${\rm N_{H,warm}: 0.7}$--1.5$\times 10^{22}$~cm$^{-2}$; Kriss et al.
1996). Mathur et al. 
(1997) investigated the possible connection between the X-ray
absorber and the strong, broad and variable UV lines (Voit,
Shull \& Begelman 1987;
Walter et al. 1990; Kolman et al. 1993; Koraktar et al. 1996), and suggested
that the absorbers are indeed one and the same outflowing system.
A quasi-simultaneous {\it Ginga} and HST observation of NGC\thinspace3516
suggested the presence of significant amount of cold photoelectric
absorption (${\rm N_{H,cold} \equiv N_H
\sim 4 \times 10^{22}}$~cm$^{-2}$) as well as
the warm absorber. The limited soft X-ray coverage of the {\it Ginga}
Large Array Counter did not allow an 
unambiguous spectral deconvolution so the
different contribution to the complex absorption could not be
separated. George et al. (1998a)
comment that the rather flat (${\rm \Gamma \simeq 1.5}$) spectral
index and the substantial amount of neutral absorption measured by
{\it Einstein}, EXOSAT and {\it Ginga} were (``undoubtedly'') due
to an incorrect modeling of ionized features with neutral gas.

The observations performed by BeppoSAX (Boella et al. 1997a)
allows for the first time the spectrum of NGC\thinspace3516 to be 
simultaneously
measured over more than three
decades in X-ray energy (0.1--200~keV),
thus ensuring the best spectral deconvolution so far.
Although the energy resolution of the BeppoSAX instruments is 
insufficient to allow high-resolution spectroscopy of the soft X-ray
features, drastic changes in the properties of the
X-ray absorber are clearly evident between the two observations
performed about four months apart.
This is the main focus of this {\it paper}, together with the presentation of
the still unpublished X-ray broadband spectrum. Sect.~2 describes the
observations and the reduction procedures. The presentation of the BeppoSAX
results is dealt with in Sects.~3 and 4 and our findings are discussed in 
Sect.~5.

\section{Observation and data reduction}

The Italian-Dutch satellite BeppoSAX carries four co-aligned Narrow Field
Instruments. Two imaging gas scintillation proportional counters:
the Low Energy Concentrator Spectrometer (LECS,
0.1--10~keV, Parmar et al. 1997) and the Medium Energy Concentrator
Spectrometer (MECS, 1.8--10.5~keV, Boella et al. 1997b). The other two
instruments have non-imaging detectors and use rocking collimators 
to monitor the background: the High Pressure
Gas Scintillator Proportional Counter (HPGSPC, 4-120~keV, Manzo et al.
1997) and the Phoswitch Detector System (PDS, 13-200~keV, Frontera et al.
1997). The HPGSPC is optimized for spectroscopy of bright sources with
good energy resolution, while the PDS possesses an unprecedented
sensitivity in its energy bandpass.

BeppoSAX observed NGC\thinspace3516 twice: on 1996 November 8 between 02:34 and
22:52 UTC, and between 1997 March 11 06:06 and March 12 01:01 UTC.
Exposure times for each instrument are shown in Tab.~\ref{tab1},
together with the observed background-subtracted count rates in the
0.1--2~keV, 2--10~keV and 15--200~keV energy ranges. Data were
\begin{table*}
\begin{footnotesize}
\caption{BeppoSAX observations log. ${\rm T_{exp}}$ and ${\rm CR}$
are the total effective exposure time and count rate in the 0.1--2~keV,
2--10~keV, and 15--200~keV energy ranges for the LECS, MECS, and PDS,
respectively.}
\label{tab1}
\vspace{0.05cm}
\begin{center}
\begin{tabular}{lcccccc}
Source & ${\rm T^{LECS}_{exp}}$ & ${\rm CR^{LECS}}$ & ${\rm T^{MECS}_{exp}}$ & ${\rm CR^{MECS}}$ & ${\rm T^{PDS}_{exp}}$ & ${\rm CR^{PDS}}$ \\
& (ks) & (s$^{-1}$) & (ks) & (s$^{-1}$) & (ks) & (s$^{-1}$) \\ 
Nov.~96  & 16.4 & $(3.40 \pm 0.16) \times 10^{-2}$ & 55.7 & $0.375 \pm 0.03$ & 41.5 & $0.78 \pm 0.05$ \\
Mar.~97  & 16.6 & $0.130 \pm 0.003$ & 45.7 & $0.697 \pm 0.004$ & 53.8 & $1.05 \pm 0.05$ \\ 
\end{tabular}
\end{center}
\end{footnotesize}
\end{table*}
telemetred in direct modes for all instruments, which provide
full information about the arrival time, energy, burst length/rise time
(RT) and, when available, position for each photon.

In this {\it paper} data from all the scientific payload BeppoSAX instruments
are presented. Standard reduction procedures and screening criteria have
been adopted to produce linearized and equalized event files. In
particular, time intervals have been excluded from the scientific product
accumulation when the angle between the pointing direction and the Earth's
limb was $< 5^{\circ}$ and the momentum associated to Geomagnetic Cutoff
Rigidity was $> 6$~GeV/c as the satellite passed through the South Atlantic
Geomagnetic Anomaly (SAGA).

The PDS data have been further screened by eliminating 5 minutes after each
SAGA passage to avoid gain instabilities due to the recovery to the
nominal voltage value after instrumental switch-on. The RT selection has
been performed using crystal temperature dependent thresholds (instead of
the fixed thresholds in the standard processing). Spectra of the imaging
instruments have been extracted from circular regions of radius 8' around
the centroid of the source for the LECS and MECS. Background
subtraction was performed using spectra extracted from blank sky
exposures in the same region as the source. The HPGSPC and PDS
background-subtracted spectra and light curves
were produced by subtraction of the ``off-'' from the
``on-source'' data. The spectra have been rebinned in order to
oversample the Full-Width Half-Maximum
of the energy resolution by a factor not greater than 3 and to have
at least 20 counts per spectral channel, to ensure the applicability
of the ${\rm \chi^2}$ test.
Uncertainties are quoted at 90\% level of confidence for one
interesting parameters (${\rm \Delta \chi^2 = 2.71}$;
Lampton et al. 1976) and energies are in the source rest frame;
$H_0 = 50$~km~s$^{-1}$~Mpc$^{-1}$
is assumed, unless otherwise specified.

\section{The NGC\thinspace3516 SPECTRAL ENERGY DISTRIBUTION}

\begin{figure}
\begin{center}
\psfig{figure=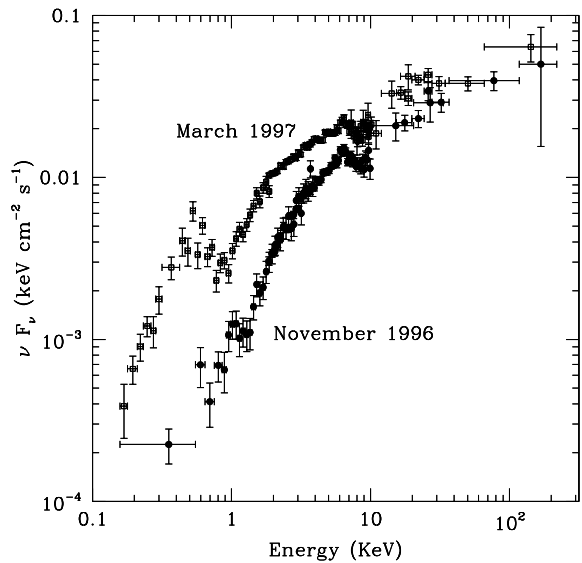,height=8.0cm,width=8.0cm,angle=0}
\end{center}
\vspace{-0.5cm}
\caption{X-ray Spectral Energy Distribution for the NGC\thinspace3516 BeppoSAX
observations of 1996 November ({\it filled circles}) and 1997 March
({\it empty squares}).}
\label{fig1}
\end{figure}

\begin{figure}
\begin{center}
\psfig{figure=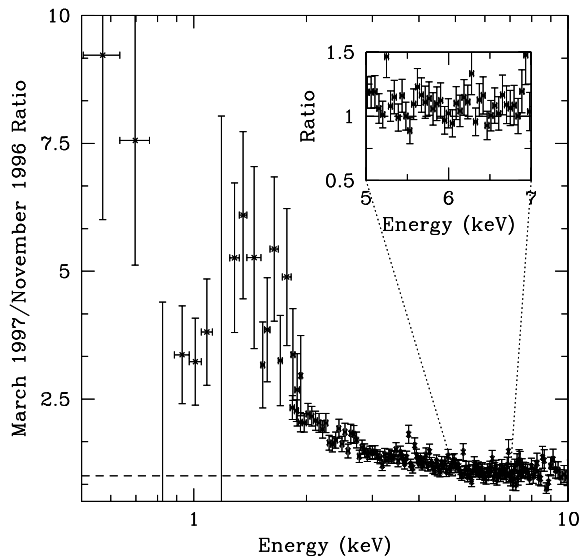,height=8.0cm,width=8.0cm,angle=0}
\end{center}
\vspace{-0.5cm}
\caption{LECS (${\rm E<2}$~keV) and MECS (${\rm E \ge 2}$~keV) spectral
ratio between the BeppoSAX NGC\thinspace3516 1997 March and 1996 November
observations. The {\it inset} shows a blow-up of the ratio in the 5--7~keV
band.}
\label{fig2}
\end{figure}

Inspection of Tab.~\ref{tab1} suggests that a dramatic spectral
change occurred between the two BeppoSAX observations.
The 0.1--2~keV:2--10~keV:15--20~keV count rate ratios were 0.04:0.48:1 and
0.12:0.67:1 in 1996 November and 1997 March, respectively, suggesting
a much softer source in the second observation. This is more quantitatively
shown in Fig.~\ref{fig1}, where the X-ray Spectral Energy
Distributions (SED) measured by BeppoSAX
are compared. The 1996 November SED was fainter
in the whole 0.1--50~keV energy range, 
and significantly harder. In contrast,
the 1997 March spectrum exhibits a deep absorption structure, starting at
$\simeq$0.7~keV. A similar
structure, if present in 1996 November, is much less pronounced.
The effect is
even more clearly displayed in Fig.~\ref{fig2}, where the 1997 March
versus 1996 November LECS/MECS spectral ratio is presented. It increases
dramatically for ${\rm E \simlt 5}$~keV. A deep depression in the
ratio between 0.7 and 1~keV suggests the presence of a localized
absorption feature in the second spectrum which is absent, or much less
pronounced, in the earlier one.

\section{Spectral analysis}

No significant spectral variability was found during either BeppoSAX
observation, {\it e.g.} between the 0.5--1.5~keV and 2--10~keV energy
ranges (in the former the contribution of the warm absorber to the flux changes
is the strongest). We therefore in this section focus on the
time-averaged spectra only, measured separately during the two observations.

The ASCA Solid Imaging Spectrometer (SIS; Gendreau 1995) has a better
energy resolution than the BeppoSAX instruments in the overlapping energy
band (0.5--10~keV). Therefore, ASCA results may represent a valuable
``pathfinder'' for a proper modeling of the BeppoSAX spectra, especially
for narrow-band features such as lines and absorption edges. NGC\thinspace3516
has been observed several times by ASCA, and the results are
discussed in various papers (Kriss et al. 1996; Reynolds 1997; George
et al. 1998a; Nandra et al. 1999). All the above papers agree on a
``baseline'' spectral model, whose main ingredients are:
a) a flat ${\rm \Gamma \simeq 1.7}$ (this results is, however,
challenged by George et al. 1998a) power-law continuum; b) a relativistic
iron K$_{\alpha}$ fluorescent emission line;
c) a warm absorber. The contribution of the cold photoelectric absorption
due to our Galaxy is generally well accounted for by a fixed column
density ${\rm N_{H,Gal} = 3 \times 10^{20}}$~cm$^{-2}$.

Time-averaged spectra were created from
cleaned event file lists, extracted from the ASCA public archive. 
The data reduction followed standard
criteria, as in Nandra et al. (1997a), except for the spectral rebinning,
which followed the recipe given in Sect.~2. We could basically reproduce
the main results presented by the above authors and we will not
include the results of this reanalysis here for
simplicity (note, however, that we use our own results when describing
the properties of the ASCA observations in
Sect.~5).

Given the much wider energy coverage of BeppoSAX,
the baseline model defined for ASCA is not appropriate.
Following the approaches of Guainazzi et al. (1999) and Perola et al.
(1999), a Compton-reflection component
(model {\tt pexrav} in {\sc Xspec}; Magdziarz \& Zdziarski 1995)
was included in the baseline model.
The relative normalization ${\rm R}$ between
the reflected and the transmitted continua (equal to 1 for reflection of an
isotropically emitted primary continuum by
a plane-parallel infinite slab) and the intrinsic cut-off energy
${\rm E_c}$ of the primary continuum were allowed to vary. 
In order not to overfit
the BeppoSAX data other reflection model parameters were held fixed
at physically meaningful values. The disk inclination was
fixed to the value derived from the long-look 1999 ASCA
observation (${\rm 35^{\circ}}$; Nandra et al. 1999),
assuming that the iron line is
produced in an X-ray illuminated relativistic disk
(model {\tt diskline} in {\sc Xspec}, Fabian et al 1989).
Solar abundances were assumed throughout.
Finally, we assumed that the iron line emitting region extends
to the innermost
stable orbit around a Schwarzschild black hole, and a radial
emissivity parameter ${\rm q = -2}$ (Nandra et al. 1997a).

We have chosen a simple parameterization of the
warm absorber with individual absorption edges,
for ease of comparison with previous works. We have then added
edges individually to the fit until statistically required
at a confidence level $>$99\%, according to the F-test.

\begin{figure}
\begin{center}
\psfig{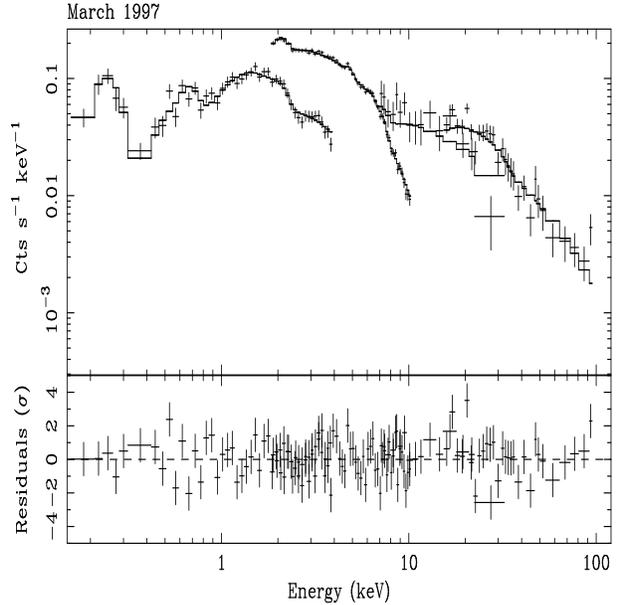}
\end{center}
\vspace{-0.5cm}
\caption{Spectra ({\it upper panel}) and residuals in units of standard
deviations ({\it lower panels}), when the BeppoSAX baseline model
is applied to
the time-averaged BeppoSAX spectra of the 1997 March observation.}
\label{fig4}
\end{figure}

This modified baseline model (``BeppoSAX baseline'' hereafter)
provides an acceptable fit to the 1997 March
BeppoSAX time-averaged spectrum with a $\chi^2$ of 328.5 for 284
degrees of freedom (dof;
see Tab.~\ref{tab2}),
\begin{table*}
\begin{footnotesize}
\caption{Best-fit parameters when the BeppoSAX
baseline model (see Sect.~4) is applied to the
1997 March time-average spectrum. The quantities with the subscript $i$
represent the photoionization absorption edges quantities}
\label{tab2}
\vspace{0.05cm}
\begin{tabular}{ccccccccccc} 
${\rm N_H}$$^a$ & $\Gamma$ & ${\rm E_{Fe}}$$^b$ & ${\rm \log (R_o)}$$^c$ & ${\rm EW_{Fe}}$$^d$ & ${\rm E_1}$ & ${\rm \tau_1}$ & ${\rm E_2}$ & ${\rm \tau_2}$ & ${\rm E_3}$ & ${\rm \tau_3}$ \\
&  & (keV) & & (eV) &  (keV) & & (keV) & & (keV) \\ 
$4 \pm 2$ & $1.60 \pm 0.03$ & $6.52 \pm^{0.13}_{0.48}$ & $0.78 \pm^{0.04}_{0.01}$ & $170 \pm^{100}_{90}$ & $0.71 \pm 0.03$ & $1.4 \pm^{0.5}_{0.7}$ & $0.91 \pm 0.07$ & $0.5 \pm^{0.7}_{0.4}$ & $7.73 \pm^{0.19}_{0.17}$ & $0.30 \pm^{0.09}_{0.08}$ \\  
\end{tabular}
\begin{tabular}{l}
$^a$in units of $10^{20}$~cm$^{-2}$
\\
$^b$rest-frame photon energy of the relativistic line
\\
$^c$logarithm of the outer radius of the iron line photon emitting region
(expressed in units of gravitational radii)
\\
$^d$iron line equivalent width
\\
\end{tabular}
\end{footnotesize}
\end{table*}
and the corresponding residuals are quite smooth
(see Fig.~\ref{fig4}). Three edges are required, whose threshold energies
are consistent, within the statistical uncertainties, with K$_{\alpha}$
photoionization from O~{\sc vii} (${\rm E_{th} = 0.737}$~keV),
O~{\sc viii} (${\rm E_{th} = 0.871}$~keV) and Fe{\sc xv}--{\sc xix}.

\begin{figure}
\begin{center}
\psfig{figure=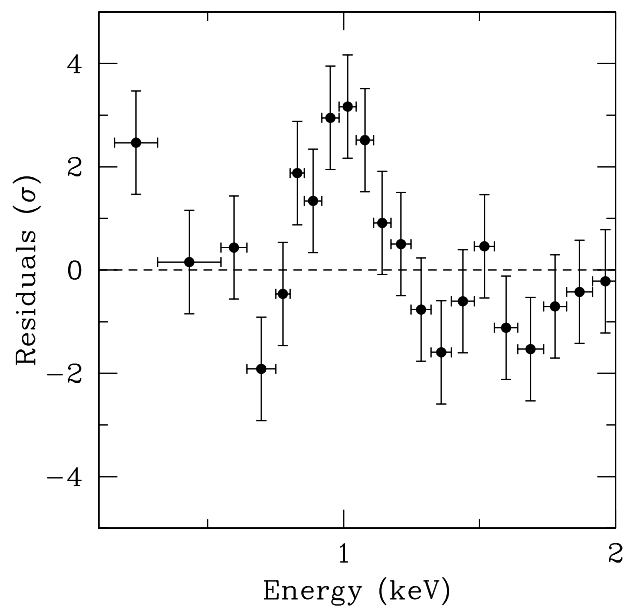,height=8.5cm,width=8.5cm}
\end{center}
\vspace{-0.5cm}
\caption{LECS residuals in units of standard deviations below 2~keV, when the
best-fit baseline model is applied to the 1996 November BeppoSAX spectrum,
once the O~{\sc vii} and O~{\sc viii} absorption edges threshold energies
are fixed at their physical values in the source rest frame.}
\label{fig5}
\end{figure}

The situation in the 1996 November spectrum is much more complex.
In Fig.~\ref{fig5}, the LECS residuals against a fit with
the BeppoSAX baseline model are shown, where the energy thresholds of 
the absorption
edges below 1~keV were held fixed at their physical values in the source
rest frame. The fit quality is poor, and line- or edge-like features
are evident. Formally, an excellent fit is obtained if the
edge energies are left free in the fit (${\rm \chi^2_{\nu} \simeq 0.92}$;
model ``B'' in Tabs.~\ref{tab3} and \ref{tab4}). In this case the best-fit
\begin{table*}
\begin{footnotesize}
\caption{Best-fit parameters and results for the continuum and iron line
components in the BeppoSAX NGC\thinspace3516 1996 November observation. 
Description
of the models: B~=~BeppoSAX baseline; B+G~=~BeppoSAX baseline with unresolved
soft X-ray emission line; B+B~=~BeppoSAX baseline plus blackbody;
PC~=~partial covering (more details in text). The values for the
absorption edges and additional soft X-ray component in models ``B'',
``B+G'' and ``B+B'' are reported in Tab.~4}
\label{tab3}
\vspace{0.05cm}
\begin{tabular}{lccccccc}  
Model & ${\rm N_H}$$^a$ & ${\rm C_f}$$^b$ & $\Gamma$ & ${\rm E_L}$$^c$ & ${\rm \log (R_o)}$$^d$ & ${\rm EW}$$^e$ & $\chi^2/$dof \\
& & (\%) & & (keV) & & (eV) & \\ 
B & $18 \pm 4$ & 100$^f$ & $1.32 \pm^{0.14}_{0.16}$ & $6.46 \pm 0.14$ & $< 1.3$ & $300 \pm 20$ & 250.5/272 \\ 
B+G & $27 \pm^{13}_{10}$ & 100$^f$ & $1.28 \pm 0.12$ & $6.4 \pm 0.2$ & $<1.1$ &$310 \pm 120$  & 257.4/273 \\ 
B+B & $140 \pm 20$ & 100$^f$ & $1.46 \pm 0.17$ & $6.4 \pm^{1.1}_{0.3}$ & $<$1.2 & $330 \pm 140$ & 246.7/273 \\
PC &  $210 \pm^{30}_{20}$ & $84 \pm 4$ & $1.53 \pm^{0.11}_{0.09}$ & $6.5 \pm 0.3$ & $> 0.84$ & $290 \pm 120$ & 240.2/271 \\ 
\end{tabular}
\begin{tabular}{l}
$^a$in units of $10^{20}$~cm$^{-2}$
\\
$^b$covering fraction of the cold photoelectric absorber
\\
$^c$rest-frame photon energy of the relativistic line
\\
$^d$logarithm of the outer radius of the iron line photon emitting region
(expressed in units of gravitational radii)
\\
$^e$iron line equivalent width
\\
$^f$fixed
\\
\end{tabular}
\end{footnotesize}
\end{table*}
\begin{table*}
\begin{footnotesize}
\caption{Best-fit parameters and results for the absorption edges
and additional soft X-ray component (rows 2 and 3 only)
in the BeppoSAX NGC\thinspace3516 1996 November observation.
The other parameters are reported in Tab.~3
The labeling of the
models is the same as in Tab.~3}
\label{tab4}
\vspace{0.05cm}
\begin{tabular}{lccccccccc}  
Model & ${\rm E_1}$ & ${\rm \tau_1}$ & ${\rm E_2}$ & ${\rm \tau_2}$ & ${\rm E_3}$ & ${\rm \tau_3}$ & ${\rm E_L}$$^b$/${\rm kT}$$^c$ & ${\rm EW_L}$$^d$ \\
& (keV) & & (keV) & & (keV) & & (keV)/(eV) & (keV) \\ 
B & $0.67 \pm^{0.10}_{0.08}$ & $2.5 \pm 0.6$ & $1.17 \pm 0.06$ &  $ 1.79 \pm^{0.19}_{0.10}$ & $7.7 \pm^{0.2}_{0.3}$ & $0.36 \pm 0.06$ & & \\
B+G & 0.737$^a$ & $< 2.6$ & 0.871$^a$ &$4.2 \pm^{0.9}_{1.5}$ & $7.7 \pm^{0.3}_{0.2}$ & $0.38 \pm^{0.12}_{0.09}$ & $1.06 \pm^{0.02}_{0.05}$ & $1.8 \pm^{1.0}_{0.8}$ \\
B+B & 0.737$^a$ & $2.3 \pm 1.2$ & 0.871$^a$  & $<$0.7 & $7.8 \pm 0.3$ & $0.30 \pm^{0.08}_{0.10}$ & $74 \pm^5_7$ & \\ 
\end{tabular}
\begin{tabular}{l}
$^a$fixed
\\
$^b$centroid energy of the soft X-ray Gaussian line
\\
$^c$temperature of the blackbody component
\\
$^d$EW of the soft X-ray Gaussian line
\end{tabular}
\end{footnotesize}
\end{table*}
values are now ${\rm E \simeq 0.67}$~keV and ${\rm E \simeq 1.17}$~keV.
Alternatively, good fits are obtained keeping the edge threshold energies
fixed at the O~{\sc vii} and O~{\sc viii} values and adding an unresolved
Gaussian emission line (model ``B+G'' in Tabs.~\ref{tab3} and \ref{tab4};
${\rm \chi^2_{\nu} \simeq 0.93}$) or a blackbody component
(model ``B+B'' in  in Tabs.~\ref{tab3} and \ref{tab4};
${\rm \chi^2_{\nu} \simeq 0.92}$). All the above models have problems,
which make them physically implausible, or not self-consistent.
We discuss the implications of these results 
in Sect.~5. In general,
we find a very hard intrinsic spectrum
(${\rm \Gamma \simeq 1.3}$) - and, correspondingly, very low values
of ${\rm E_c}$ (${\rm \simeq 30}$~keV) and R (${\rm \simlt 0.3}$) -
and a surprisingly
large soft X-ray line with an EW of ${\rm \simeq 1.8}$~keV,
or blackbody fluxes. Moreover, all models require a substantial ``cold''
photoelectric column of 0.2--1.1$\times 10^{22}$~cm$^{-2}$.

\begin{figure}
\begin{center}
\psfig{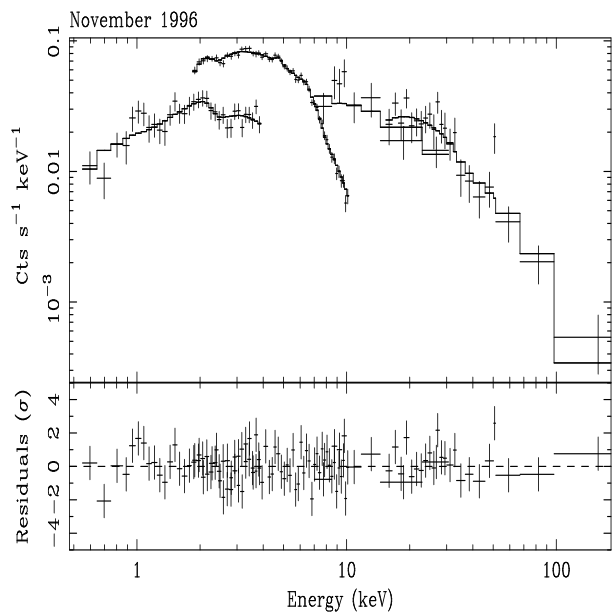}
\end{center}
\vspace{-0.5cm}
\caption{Spectra ({\it upper panel}) and residuals in units of standard
deviations ({\it lower panel}), when the model with partial covering
neutral photoelectric absorption is applied to the
the time-averaged BeppoSAX spectra of the 1996 November observation.}
\label{fig6}
\end{figure}

\begin{figure}
\begin{center}
\psfig{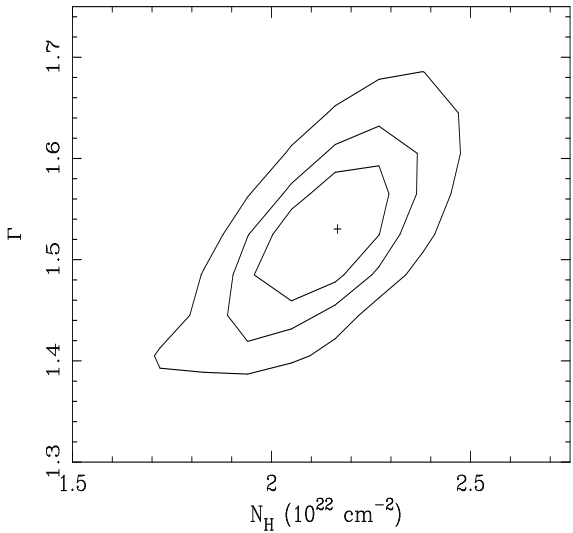}
\end{center}
\vspace{-0.5cm}
\caption{Column density versus spectral index iso-${\rm \chi^2}$ contour
plot for the ``PC'' model applied to the November 1996 BeppoSAX
spectra. The contours represent the 68\%, 90\% and 99\% confidence
levels for two interesting parameters}
\label{fig9}
\end{figure}

A much simpler, elegant and self-consistent solution is 
to substitute the warm with a cold absorber, and allow its 
covering fraction to be a free parameter. The fit quality is again excellent
(${\rm \chi^2_{\nu} = 0.92}$) and the residuals unstructured
(see Fig.~\ref{fig6}). In this framework, the properties of the intrinsic
nuclear spectrum are very similar during the two BeppoSAX observations
(cf. Tab.~\ref{tab2} and Tab.~\ref{tab3}).
The inferred {\it cold} column density (${\rm N_H
\simeq 2 \times 10^{22}}$~cm$^{-2}$), is more than two orders of magnitude 
higher than measured in any other ROSAT,
ASCA or BeppoSAX observation,
and its covering fraction $>$80\%. If
absorption edges from O~{\sc vii}, O~{\sc viii}
or ionized iron (with ${\rm E_{th} = 7.7}$~keV) are added to the
model, only upper limits
are obtained with optical depths, of 1.0, 0.2 and 0.4, respectively.
The 90\% upper limit on the ionization parameter ${\rm \xi}$\footnote{${\rm
\xi \equiv L/n R^2}$, where ${\rm L}$ is the integrated luminosity of
the nuclear source in the 5~eV--300~keV energy range (Done et al.
1992), ${\rm n}$ is the plasma electron density and ${\rm R}$ its
distance from the illuminating source} of the cold absorbing matter
is 1.6, assuming ${\rm T = 2 \times
10^5}$~K. In Fig.~\ref{fig9} we show the
column density versus spectral index iso-${\rm \chi^2}$ contour plot.
At the 99\% confidence level for two interesting parameters, ${\rm N_H}$
is comprised between 1.4 and 2.4$\times 10^{22}$~cm$^{-2}$.
We consider
the ``partial covering'' scenario the most plausible model for
the 1996 November BeppoSAX time-averaged spectrum, and will discuss
its properties in the remainder of this section.

We note that the ``warm'' and ``neutral partial covering'' scenarios for
the absorber are not mutually interchangeable.
The latter scenario yields a ${\rm \chi^2 = 364.1/288}$~dof if applied
to the 1997 March data. This conclusion applies {\it a fortiori} to the ASCA
observations. The ${\rm \chi^2}$ is 573.1 for 532~dof and 640.6 for 539~dof
for the former scenario in the 1994 April and 1995 March observations
(our reanalysis), whereas it is 1647/547~dof and 950.6/544~dof in the latter,
respectively.

The BeppoSAX results confirm that the spectral index of the primary
continuum in NGC\thinspace3516 (${\rm \Gamma \simeq 1.5}$)
is flatter than typically observed in
Seyfert galaxies (Nandra et al. 1997), albeit not extreme.
In Tab.~\ref{tab6} the best fit values for ${\rm R}$ and ${\rm E_c}$
\begin{table}
\begin{footnotesize}
\caption{Best-fit Compton-reflection versus primary continuum relative
normalization (${\rm R}$) and cut-off energy (${\rm E_c}$) in the
two BeppoSAX observations.}
\label{tab6}
\vspace{0.05cm}
\begin{center}
\begin{tabular}{lcc} 
Date & R & ${\rm E_c}$ \\ 
& & (keV) \\ 
1996 November & $0.7 \pm^{0.5}_{0.4}$ & $140 \pm^{220}_{60}$ \\
1997 March & $0.7 \pm^{0.4}_{0.3}$ & $120 \pm^{80}_{40}$ \\  
\end{tabular}
\end{center}
\end{footnotesize}
\end{table}
are shown. Despite a factor
$>$1.5 increase of the 0.5--10~keV flux between the two BeppoSAX
observations, both quantities remain constant, in agreement with
the disk reflection paradigm.

The best-fit models correspond in the 1996 November
(1997 April) observation to observed fluxes of 0.27 (1.36) and
2.50 (4.41) $\times 10^{-11}$~erg~cm$^{-2}$~s$^{-1}$ in the
0.5--2~keV and 2--10~keV energy bands, respectively . These values 
correspond to unabsorbed rest-frame luminosities of 0.43 (0.71) and
1.01 (1.56)$\times 10^{43}$~erg~s$^{-1}$, respectively.

\section{Discussion}

We compare the time-averaged spectra of NGC\thinspace3516
measured by BeppoSAX during two observations separated by four months.
The earlier spectrum is weaker and harder
in the whole 0.1--50~keV energy range. In addition,
the deep absorption edges from
ionized oxygen, which have been observed
by ROSAT (Mathur et al. 1997) and ASCA (Kriss et al. 1996;
Reynolds 1997; George et al. 1998a) are much less pronounced.
These remarkable
spectral changes could in principle
reflect a dramatic modification either of the condition
of the cold/warm photoelectric absorption, or of the
mechanism producing the primary nuclear continuum.

We propose that these changes are most easily
interpreted as due to the onset of substantial cold
absorption (${\rm
N_H \simeq 2 \times 10^{22}}$~cm$^{-2}$) along the line of sight to
the nucleus during the earlier observation.
The properties of the primary continuum and of the
Compton reflection components do not appear to have substantially changed
between the two observations, despite an overall increase of the 2--10~keV
flux by a factor about 1.5. 

\subsection{NGC3516: a low-luminosity BAL quasar?}

This is not the first claim that the nucleus of
NGC\thinspace3516 is seen through a
substantial ``cold'' column density. The same
conclusion was reached by
Kruper et al. (1990), on the basis of 
{\it Einstein} MPC observations
and by Ghosh \& Soundararajaperumal (1991) using
EXOSAT. A 1989 {\it Ginga} spectrum also required a
column density of ${\rm N_H \simeq 4 \times 10^{22}}$~cm$^{-2}$
(Kolman et al. 1993). However, in all subsequent ROSAT and
ASCA observations, NGC\thinspace3516 turned out to be a ``standard'' warm
absorbed Seyfert~1. This casts doubt on the previous interpretations,
which were obtained using less sensitive instruments.
The BeppoSAX finding is the first confirmation with
moderate resolution detectors
that NGC\thinspace3516 indeed undergoes phases with large amounts
of absorption by cold material.

It is difficult, currently, to
unambiguously characterize the physics and/or geometry underlying these
changes. We investigate two main options in the following:
{\it i)}: the cold absorber constitutes a different absorbing system from
the warm absorber; {\it ii)} the observed variability reflects changes
in the physical properties of the warm absorber itself.

The former scenario can be simply described as the interposition
of ``bricks'' of matter, with ${\rm N_H \sim 10^{22}}$--$10^{23}$~cm$^{-2}$,
along the line of sight to the nucleus. The available observations have a too 
sparse a pattern to provide any strong quantitative constraints on the nature
of these ``bricks''. It is likely that there are a small number of
big clouds, because it is difficult
to reproduce simultaneously the apparent ``on-off''
pattern of high absorption occurrences
(from null to $\simgt 10^{22}$~cm$^{-2}$) and the lack of significant
spectral variability within an observation of typical duration a few
days with a large number of randomly
distributed clouds.
The properties of the absorbing cloud(s) can be constrained, using
the limits on the variability timescales available from the BeppoSAX
observations ({\it i.e.}: minimum of 1~day - the length of the
November 1996 observation - and maximum of 4~months - the distance
between the two observations), and the upper limit on the ionization
parameter of the absorbing matter (${\rm \xi < 1.6}$). Assuming
a single spherical cloud of radius ${\rm r}$ rotating with
Keplerian velocity ${\rm v_K}$
around a black hole of mass ${\rm M_6 = M_{BH}/10^6 M_{\odot}}$:
$$
v_K = ( \frac{G M_{BH}}{R} )^{1/2} = ( \frac{\xi N_H}{L r} )^{1/4} (G M_{BH})^{1/2}
$$
The ``disappearance'' of the cold absorber in the April 1997
observation implies ${\rm r/v_K \simlt 10^7}$~s, or:
$$
r^{5/4} < 4 \times 10^{26} N_H^{1/4} L^{-1/4} (G M_6)^{1/2} \simeq 6 \times 10^{17} cm^{5/4}
$$
or ${\rm r < 6 \times 10^{14}}$~cm (${\rm L = 6 \times 10^{43}}$~erg~s$^{-1}$
in the November 1996 observation). The lower limit on the
distance between the absorbing cloud and the nuclear illuminating
source can be expressed
(again through the ionization parameter definition)
as ${\rm R > 2 \times 10^{17} n^{-1/2}_9}$,
(${\rm n_9}$ is the electron density in units of
${\rm 10^9}$~cm$^{-3}$). Given the above constraints, it
is interesting to test the hypothesis whether the absorbing
matter is associated with the Broad Line Regions (BLR).
We assume the measured FWHM of the H$_{\beta}$ line
(4000~km~s$^{-1}$; Wanders et al. 1993) as the velocity
of the cloud. The size of the cloud is therefore constrained
in the range ${\rm 4 \times 10^{13} \simlt r \simlt 4 \times
10^{15}}$~cm, corresponding (through ${\rm N_H \equiv nr}$)
to a range in electron densities: ${\rm 5 \times 10^6 \simlt
n < 5 \times 10^8}$~cm$^{-3}$. This range is only marginally
consistent with the current estimates of the characteristic densities
in BLR (${\rm n \sim 10^9}$--$10^{12}$~cm$^{-3}$; see {\it e.g.} Krolik et
al. 1991). Moreover, our lower limit on ${\rm R}$
is not consistent with the reverberation
mapping measurements
of the BLR size ($\sim 3 \times 10^{16}$~cm;
Wanders et al. 1993), unless {\it high} densities
are assumed.

The Broad Absorption Lines quasars (BALs) are a class of quasars,
characterized by strong high-ionization absorbing features. The UV
absorbing system in NGC\thinspace3516 (Voit et al. 1987; Walter et al. 1990)
strongly resembles those of BALs. There are, however, significant differences,
namely the lower EW
($\sim$40\AA \ in BALs versus $\simlt$10\AA \ in NGC\thinspace3516),
the range in velocity (up to 10000~km~s$^{-1}$ versus 3000~km~s$^{-1}$)
and the variability timescales (years versus weeks/months). Kolman
et al. (1993) were the first to speculate that NGC\thinspace3516 may be a 
low-redshift
counterpart of the BALs, and its peculiar properties may be mainly due to the
weaker nuclear engine energy output. The variability timescale of
the UV absorption features is of the same order as the
(most likely) occurrence of the  high X-ray absorption states.
Intriguingly, recent studies of a sample of BAL quasars
using ROSAT, ASCA and BeppoSAX (Gallagher et al. 1999; Brandt et al.
1999) discovered that they are considerably weaker X-ray sources than
expected from their optical fluxes, and column densities as high
as several $10^{23}$~cm$^{-2}$ are not uncommon. We can 
extend the original speculation further and propose that
the X-ray highly absorbed states in NGC\thinspace3516 
are connected to the onset of
UV absorption features. In the cases where simultaneous UV/X-ray
data are available, this connection is indeed present (Kolman et al.
1993; Koraktar et al. 1996; Kriss et al. 1996).
Weymann et al. (1991) suggested that the clouds responsible for
the BAL phenomenon are ablated from the rim of the molecular torus
surrounding the nuclear region of the AGN. Intriguingly,
the velocity field of the circum-nuclear emission-line regions
suggests that our line of sight towards the NGC\thinspace3516
nucleus indeed grazes the putative torus (Goad \& Gallagher 1987).
However
there are still theoretical difficulties in encompassing the cold X-ray
and UV absorption features in one and the same absorbing system.
As pointed out by Kolman et al.
(1993), it is difficult to obtain the observed width of the C~{\sc iv}
absorption feature with a column density of the order of
${\rm 10^{22}}$~cm$^{-2}$. 

\subsection{A ``non-standard'' warm absorber?}

\begin{figure}
\begin{center}
\psfig{figure=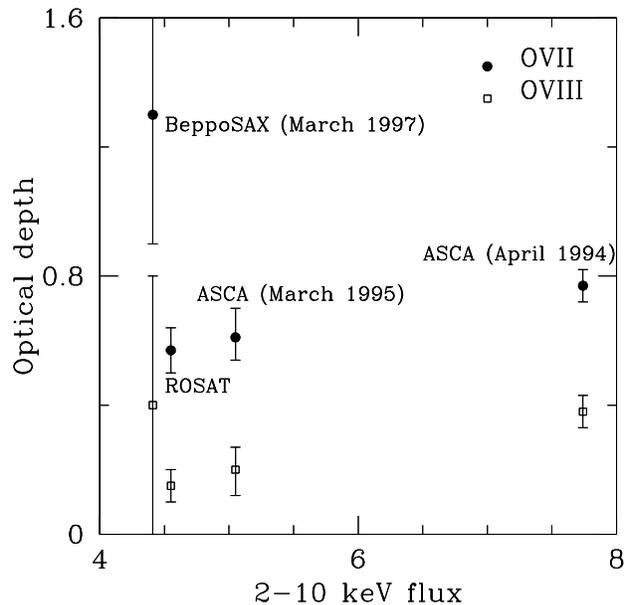,height=8.5cm,width=8.5cm}
\end{center}
\vspace{-0.5cm}
\caption{Best-fit absorption edges optical depths
against 2--10~keV flux in ROSAT (Mathur et
al. 1997), ASCA (our reanalysis) and BeppoSAX
1996 April (this {\it paper}) time-averaged spectra for O~{\sc vii}
({\it filled circles}) and O~{\sc viii} ({\it empty squares}).
The flux for the ROSAT observation
is extrapolated from the best-fit warm absorber model of Mathur et
al. (1997). In all cases the baseline model defined in Sect.~4 is 
used (Compton reflection included
for BeppoSAX data only), with the threshold energies of the edges
fixed at their physical values in the source rest frame.}
\label{fig8}
\end{figure}

Alternatively, these intervals of apparent high ``cold'' absorption could
be due to a change in the physical properties (namely the ionization
structure) of the warm absorber itself. A comparison of the
measured O~{\sc vii} and O~{\sc viii} absorption edge optical depths
with the 2--10~keV flux before BeppoSAX shows, if any,
a slight trend for both edges to become deeper with increasing flux
(Fig.~\ref{fig8}), which would exclude simple models, where the
equilibrium ionization structure of the absorber is mainly driven
by the average ionizing continuum.
Mathur et al. (1997) developed a dynamical
model to explain the changing features of the UV and X-ray absorber
(``XUV system''). According to their model, the XUV system
is outflowing with a radial velocity of $\simeq$500~km~s$^{-1}$.
The disappearance of the
UV absorption lines is due to an increase of the
ionization parameter caused by the decreasing density in the
expanding matter. They predicted a progressive weakening of the
O~{\sc vii} edge in comparison to the O~{\sc viii}. Ultimately,
the absorber would become totally transparent to X-rays. This
prediction is not fulfilled by the BeppoSAX observations presented
here. The 1997 April spectrum is well fit within the
framework of a ``standard'' warm absorber scenario, and the O~{\sc vii}
absorption edge optical depth is almost double the maximum observed
by ROSAT (Mathur et al. 1997) and ASCA (Kriss et al. 1996; Reynolds 1997;
our reanalysis of archival data; see Fig.~\ref{fig8}). A new absorbing
system must have been produced in the meanwhile.

Another question is whether differing
conditions of the warm absorber can {\it mimic} a substantial neutral
column density in instruments with moderate energy resolution such as
those on BeppoSAX.
Actually,
the 1996 November spectrum can be formally well fit within a
warm absorber framework, where the opacity is described by two
absorption edges of ${\rm \simeq
0.67}$ and ${\rm \simeq 1.17}$~keV. 
The recent discovery of the so-called ``1~keV'' warm
absorber features in the X-ray spectrum of some Narrow Line Seyfert 1
Galaxies (Leighly et al. 1997) has raised new interest and theoretical
speculations on this subject. Nicastro et al.
(1999a, 1999b) demonstrate that strong absorption features in the range
1--3~keV can be produced under two different physical
conditions: a) if the ionizing continuum if steeper than observed
in ``broad'' Seyfert 1 galaxies; b) or - and this is the scenario
which may be relevant to our case - if collisional
ionization dominates. In the
latter scenario, the highest source of opacity is provided by ionized
species of neon and iron (L-transitions).
However, this explanation presents several
problems. 
The first
is that the 0.67~keV feature has an energy appropriate
to O~{\sc vi} (Lithium-like oxygen), which only dominates the ionic
distribution for a narrow range of ionizing fluxes.
Similar considerations apply if the feature is instead identified
with N~{\sc vii}.
However, the observed feature is not formally inconsistent, within the
statistical uncertainties, with the more common
K$_{\alpha}$ photoionization from O~{\sc vii}.
The 1.17~keV feature can be associated with
either Ne~{\sc ix} or Fe~{\sc xvii}--{\sc xviii}. A range of electron
temperatures around a few $10^6$~K exist, where these species have
simultaneously the highest relative fractional abundance
in a collisionally ionized plasma (Nicastro et al. 1999a).
Such high temperatures would be associated with a ionization distribution
of oxygen where the contribution of O~{\sc viii} is important, if not dominant,
making the identification of the lower energy edge even more problematic.
However, Nicastro et al. (1999a) show that the simultaneous presence
of O~{\sc vii} and ``1-keV'' features can be achieved with mixed
collisional-photoionization models, which also predict
a decoupling of the ionization parameter of the gas from the
ionizing continuum flux.
The main obstacle to the viability of this model remains, however, the
extremely flat intrinsic power-law spectrum that requires
(${\rm \Gamma \simeq 1.3}$) - and the correspondingly low cutoff energy
(${\rm E_{c} \simeq 30}$~keV) -,
which has never been observed in NGC\thinspace3516
or other Seyfert~1 galaxies (Nandra \& Pounds 1994; Nandra et al. 1997a).
We therefore consider this a very unlikely possibility.

\subsection{The iron edge problem}

The 1997 March BeppoSAX observation baseline best-fit model requires
an absorption edge, whose threshold energy (${\rm 7.7}$~keV) corresponds to
K$_{\alpha}$ photoionization
from highly ionized iron (Fe{\sc xv}--{\sc xix}). It is very unlikely that
the oxygen and the iron absorption features originate in the same
warm absorber. If we assume the photoionization cross sections of
Verner \& Yakovlev (1995) and the abundances of Anders \& Grevesse
(1989), the hydrogen equivalent column densities for oxygen and
iron differ by more than one order of magnitude: ${\rm
N^H_0 \sim 1.2 \times 10^{21}}$~cm$^{-2}$, ${\rm N^H_{Fe}
\sim 3 \times 10^{22}}$~cm$^{-2}$. The iron
feature is at 20\% of the best-fit continuum/data ratio,
then significantly deeper than any reported calibration uncertainties
in the MECS at those energies ($\simeq$5\%; Fiore et al. 1999). Moreover,
analogous features are detected in the publicly available ASCA spectra.
In Tab.~\ref{tab9} we report the best-fit threshold energy and optical depth
\begin{table}
\begin{footnotesize}
\caption{Best-fit parameters for the ionized iron absorption edge in the ASCA and March 1997 BeppoSAX observations. The ${\rm \Delta \chi^2}$ refers to the addition of this feature to the model (corresponding to 2 less dofs). The best-fit model is the baseline (without Compton reflection for the ASCA observations)}
\label{tab9}
\vspace{0.05cm}
\begin{center}
\begin{tabular}{lccc} 
Date & ${\rm E_{th}}$ & ${\rm \tau}$ & ${\rm \Delta \chi^2}$ \\ 
1994 April$^a$ & $7.75 \pm 0.17$ & $0.27 \pm^{0.07}_{0.06}$ & 57 \\
1995 March$^a$ & $7.7 \pm 0.3$ & $0.27 \pm^{0.10}_{0.12}$ & 8.3 \\
1997 March$^b$ & $7.73 \pm^{0.19}_{0.17}$ & $0.30 \pm^{0.09}_{0.08}$ & 37 \\ 
\end{tabular}
\end{center}

\noindent
$^a$ASCA

\noindent
$^b$BeppoSAX

\end{footnotesize}
\end{table}
of this feature in the ASCA observations, when the baseline model
(without Compton reflection) is applied (the 1997 March
BeppoSAX measurement is also reported for ease of comparison).
The ASCA and BeppoSAX measures are very consistent. We note in passing
the the upper limit on the optical depth of a 7.7~keV absorption edge (0.4)
in the BeppoSAX 1996 September observation is also consistent with
these values.
In principle, a strong iron overabundance in the absorbing medium could
reconcile the discrepancy. A contribution to the iron
absorption edge could come from the opacity in a partly ionized reflector.
However, the latter hypothesis is contradicted by the fact the the iron
line measured by ASCA (Nandra et al. 1999) is inconsistent with being
produced by fluorescence of significantly ionized iron.
Alternatively, the iron edge could originate in a second
warm absorber, more ionized than that responsible for the oxygen
features. We have investigated this possibility, fitting the
April 1997 BeppoSAX spectrum with a double ionized absorber
model.
We have assumed two scenarios: either the primary continuum
reaches us through a single optical path, along which the two
absorbing screens are located; or it follows two different
optical paths, and one absorbing screen is located along each
of them. The {\sc Xspec} implementation {\tt absori} for the
warm absorber has been used in these fits.
The best-fit parameters and results are summarized in
Tab.~\ref{tab10}. The quality of these fits is comparable with
\begin{table*}
\begin{footnotesize}
\caption{Best-fit parameters and results when the double ionized absorption
model is applied to the BeppoSAX April 1997 observation. The rows
correspond to either one of the following scenarios:
 a model where the absorbing screens are located along
the same optical path (row~1); the absorbing screened are located
along different optical paths
(row~2)}
\label{tab10}
\vspace{0.05cm}
\begin{center}
\begin{tabular}{lllllllllll}
\multicolumn{3}{l}{First absorber} & \multicolumn{3}{l}{Second absorber} & \multicolumn{2}{l}{Continuum} & \multicolumn{2}{l}{Iron line} \\
${\rm N_{H,1}}$ & ${\rm \log (\xi_1)}$ & ${\rm \log (T_1)}$ & ${\rm N_{H,2}}$ & ${\rm \log (\xi_2)}$ & ${\rm \log (T_2)}$ & ${\Gamma}$ & ${\rm R}$ & ${\rm E}$ & ${\rm EW}$ & ${\rm \chi^2/}$~dof \\
(${\rm 10^{22}}$~cm$^{-2}$) & & & (${\rm 10^{22}}$~cm$^{-2}$) & & & & & (keV) & (eV) & \\
$1.17 \pm^{0.17}_{0.13}$ & $0.2 \pm^{0.2}_{0.7}$ & $6.7 \pm 0.6$ & $9 \pm^6_4$ & $2.6 \pm^{0.5}_{1.0}$ & $\equiv T_1$ & $2.06 \pm^{0.08}_{0.05}$ & $2.1 \pm^{1.0}_{0.6}$ & $6.3 \pm 0.2$ & $100 \pm^{90}_{70}$ & 334.8/287 \\
$0.57 \pm^{0.08}_{0.05}$ & $<0.17$ & $5.7 \pm^{1.5}_{0.4}$ & $4.6 \pm^{2.4}_{0.8}$ & $1.9 \pm^{2.1}_{1.3}$ & $< 5.9$ & $2.15 \pm^{0.13}_{0.06}$ & $3.0 \pm^{1.3}_{0.4}$ & $6.3 \pm 0.2$ & $60 \pm^{40}_{30}$ & 342.4/286 \\ 
\end{tabular}
\end{center}

\end{footnotesize}
\end{table*}
the phenomenological fits in Sect.~4. However, they
suffer a serious inconsistency. They require a steep
intrinsic spectrum (${\rm \Gamma \simeq 2.1}$), and a correspondingly
larger amount of reflection (${\rm R = 2}$--3). On the other
hand, the iron line
EW is very small ($\simeq 60$--100~eV). This is inconsistent
with the disk paradigm (Matt et al. 1992), unless the line is
significantly ionized (Matt et al. 1993; ${\rm \dot{Z}}$ycki et al. 1994).
Again, this hypothesis is, however, not supported by the observational
evidence available so far.

\subsection{NGC3516 in hard X-rays: a reassuring ``standard'' Seyfert 1}

The other X-ray spectral features of NGC3516 are not uncommon
among Seyfert 1 galaxies. The intrinsic spectral index measured
by BeppoSAX is flat (${\rm \Gamma \simeq 1.6}$), but not
extreme. The Compton reflection is slightly smaller, but not inconsistent,
with a that expected from a plane-parallel slab subtending a
2$\pi$ angle from the nucleus ({\it i.e.} an accretion disk).
The iron line EW is consistent,
within the statistical uncertainties, to having an origin in the same
accretion disk [this would {\it not} be the case for the line EW measured
by ASCA (Nandra et al. 1997b; Nandra et al. 1999), suggesting long-term
changes in the relative weights of the reprocessed components].
It is worth noting 
the constancy of the amount of Compton reflection
between the two BeppoSAX observations, despite a 1.5 variation in the
2--10~keV luminosity. This allows
an upper limit on the size of the reflecting region of about
0.1~pc to be set. It is likely to be smaller than this. The study
of the intra-observation ({\it i.e.} intra-day) variability pattern,
is consistent with a response of the Compton reflection on
timescales as short as a few thousand seconds.
Similar conclusions were reached by Nandra et al. (1997b) from their
study of the iron line flux variability in the ASCA observations.\\

After the present work was completed, we became aware of a paper
discussing the same set of data by Costantini et al. (2000).
Although their main conclusions coincide with ours, they report the
detection of the ionized iron absorption edge also in the November
1996 observation (due to the different parameterization used,
it is not straightforward to compare their detection with our
upper limit). They attribute it to a ``highly photoionized
component [..], mainly visible when our view of the primary
X-rays is at least partially covered by a large amount of
neutral absorber''. Moreover, their fits yield a steeper intrinsic
spectral index (${\rm \Gamma = 2}$), with a correspondingly larger
amount of reflection (${\rm R \simeq 1.2}$--1.4).

\section*{Acknowledgments}

The BeppoSAX satellite is a joint Italian-Dutch programme.
M. Guainazzi acknowledges an ESA Fellowship
and W. Marshall the ESA stagiaire programme. 
This research has made use of data obtained through the High Energy 
Astrophysics Science Archive Research Center Online Service, provided 
by the NASA/Goddard Space Flight Center.

\end{document}